*Original Article*

# Mining Privacy-Preserving Association Rules based on Parallel Processing in Cloud Computing

Dhinakaran D[1], Joe Prathap P. M[2], Selvaraj D[3], Arul Kumar D[4], Murugeshwari B[5]

[1]*Research Scholar, Department of Information and Communication Engineering, Anna University, Tamilnadu, India*
[2]*Professor, Department of Information Technology, R.M.D Engineering College, Tamilnadu, India*
[3]*Professor, Department of Electronics and Communication Engineering, Panimalar Engineering College, Tamilnadu, India*
[4]*Associate Professor, Department of Electronics and Communication Engineering, Panimalar Institute of Technology, Tamilnadu, India*
[5]*Professor, Department of Computer Science and Engineering, Velammal Engineering College, Tamilnadu, India*

**Abstract** - *With the onset of the Information Era and the rapid growth of information technology, ample space for processing and extracting data has opened up. However, privacy concerns may stifle expansion throughout this area. The challenge of reliable mining techniques when transactions disperse across sources is addressed in this study. This work looks at the prospect of creating a new set of three algorithms that can obtain maximum privacy, data utility, and time savings while doing so. This paper proposes a unique double encryption and Transaction Splitter approach to alter the database to optimize the data utility and confidentiality tradeoff in the preparation phase. This paper presents a customized apriori approach for the mining process, which does not examine the entire database to estimate the support for each attribute. Existing distributed data solutions have a high encryption complexity and an insufficient specification of many participants' properties. Proposed solutions provide increased privacy protection against a variety of attack models. Furthermore, in terms of communication cycles and processing complexity, it is much simpler and quicker. Proposed work tests on top of a real-world transaction database demonstrate that the aim of the proposed method is realistic.*

## I. INTRODUCTION

The processing of enormous volumes of data into valuable patterns and rules is known as data mining. The data mining technique has been increasingly explored and utilized in numerous scientific and commercial fields because it extracts meaningful knowledge from vast amounts of data. The development of data mining techniques has significantly impacted a wide range of applications. ARM is an integral approach to identify the underlying association among items through massive data, exposing latent association patterns, and subsequently aiding in economic operations and management information systems. In high transaction databases, frequent itemset mining and ARM are two extensively utilized data processing approaches for uncovering often co-occurring collected data and intriguing association links among datasets, correspondingly [1]. These ARM methods have historically been performed over a compressed format, while all information collected into a centralized location and techniques run against specific data. Since there is no completely trustworthy third party, privacy risks arise. Service providers are frequently honest and inquisitive, wanting to learn more about users and hence open to misuse. To solve this difficulty, recommend a quasi third party. The most important focal point of this paper is the extraction of frequent patterns in dispersed collections using a semi-trusted intermediary service. Neither server nor the participants have access to the private transactions of other parties. Unofficially, the aim denotes a secure multi-party computational issue [2].

The concerns posed by the ARMS technique have lately been analyzed in security and privacy considerations. As a result, people's privacy is violated. Frequent itemset mining (FIM) can reveal prevalent itemsets and associated possibly relevant relationships from a transaction dataset [3]. After obtaining a large number of itemsets, mining association rules get easy. However, transferring the unprocessed data straight to the cloud service provider (CSP) is risky because CSP may be interested in sensitive transactions. When extracting private information, the security level must always be carefully evaluated. The exploitation of this technology has the potential to expose the data owner's perceptive information toward others.

The purpose of the ARM is to reveal frequent itemsets that frequently appear in transactional data. Before centralized mining, there was much concentration. The issue has a highest exceedingly terrible uncertainty of significant worst-case complexity, a characteristic that drives businesses to outsource mining to a cloud that has developed effective, profitable, and customized solutions. In addition to the mining cost reduction, the data owner intends to outsource the data mining task. First and foremost, it necessitates negligible computational resources because forcing the owner to create and transfer transactions with the miner [4]. This makes the outsourced approach more appealing to applications where data owners generate transactions and comprise limited resources to handle them. Second, assuming that the proprietor's aim has a variety of transaction generating sources, such as a network of business areas that produce transactions or transactions in various locations—submitting all of these transactions to a singular service provider for mining purposes [5]. The service provider could handle association regulations unique to particular businesses or apply to the entire organization. As a result, the expense of transferring transactions between two parties and doing global processing in a distributed way is reduced.

The CSP, on the other hand, becomes a single point of security attack. If a third-party service provider (SP) is untrustworthy, he should be barred from obtaining the basic information because the information is critical to the company. Similarly, regardless of whether the items are public or not, the newly formed association rules remain the owner's private property, and they are recommended to be recognized only by the data owner. As a result, guaranteeing confidentiality to both original data and creating frequent patterns by the SP is critical in sequential pattern mining task outsourcing. Various approaches really can assure the security of sensitive data. The first is to use cryptographic primitives to convert the data from one format.

On the other hand, the use of cryptography allows the precise rules to be reconstructed. Another methodology is to split data horizontally or vertically and transmit it to multiple servers so that no single outsource server gets the entire pattern because data is spread among the parties. Here, this paper proposes evaluating suitable cryptographic primitives for outsourcing ARM to incorporate both strategies and also keep increasing safety by creating cryptosystem to data [6].

Because of its exceptional capabilities for collecting, analyzing, and manipulating large amounts of data, cloud computing has become a novel diagnostic energy source. Cloud computing has a great deal of potential in providing reliable processing capacity for the collective monitoring of flexible sources [7]. Customers must encrypt their personally identifiable information before outsourcing to preserve it from illegitimate usage; however, conducting operations on this encrypted data is challenging for cloud servers. To maintain privacy when outsourcing frequent pattern extraction, the data possessor has to convert the raw data so that the CSP cannot infer or derive additional content than what the data owner has contributed, as shown in Fig. 1. Its familiarity of frequent itemsets in addition to their support as of prior practice be supposed to not consent to this extraction to lead to a breach of privacy. For example, because the top frequent things are usually unique in that area, the data miner possibly will be capable of identifying them with no trouble.

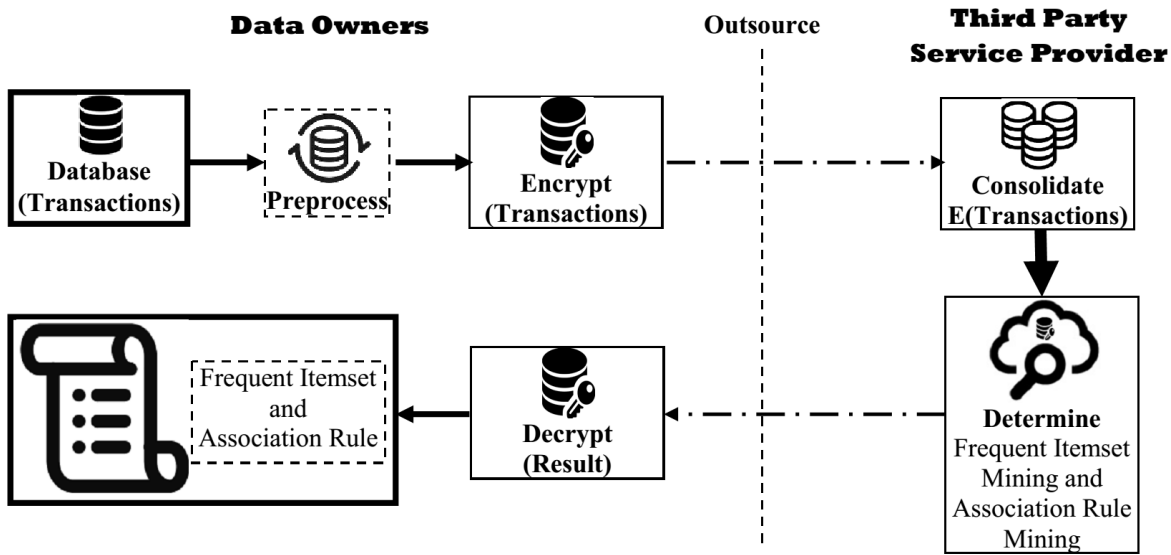

**Fig. 1 A common design for outsourced data Mining**

The work aims to improve the ARM and frequent itemset mining outcomes regarding communication complexity, computational complexity, and cost associated. With double encryption, to process frequent itemset mining and ARM for vital privacy needs. This work provides an efficient Transaction Splitter technique and a customized apriori algorithm approach to tackle this issue for horizontally separated datasets that perform privacy-preserving frequent pattern mining to keep a strategic distance from exposure of data owner's sensitive data. By looking into the trade-offs between usability and confidentiality that emerge when cloud services collect, explore, and share data, and by constructing techniques to help them manage such transfer. This paper utilizes a twofold encryption algorithm to protect data owners' original data. Through parallelism, the proposed approach improves the effectiveness of the work. Proposed technology ensures data security throughout processing and transmission and no loss of data or utility loss. The following is how the rest of the paper is ordered: In the second section, provides related work regarding PPDM. The third section explains how to put the strategy into action. The details of the experiment are found in Section 4. Section 5 ends with a conclusion and future investigation.

## II. RELATED WORK

This section summarizes most of the significant research work executed during the last couple of decades. Among the most commonly utilized PPDM on outsourced cloud data is the randomization-based method. Data perturbation applies noise to unprocessed data to safeguard private information, making mining more complex and can have unpredictably negative repercussions for mining effectiveness. Differential privacy mitigation computations can stave off contextual interpretation assaults; however, the mining outcomes are unreliable. Furthermore, due to the combinatorial property of differential privacy, this solution no more than can consent to a limited number of requests per user. In contrast to the previous method, the cryptography-based technique uses cryptographic data primitives to safeguard content in an immersive processing paradigm. It facilitates data mining in a distributed system. This strategy also addresses the issue of data loss and low down consistency in mining outcome, however, in addition, establishes a distinct sanctuary framework.

Despite the attackers' important contextual information, Qilong et al. [8] focused on maintaining differential privacy in data centers without disclosing actual transactions, using an intermediary server for data integration without assuming that it is secure. Their solutions provide increased privacy protection against a variety of attack models. Furthermore, it is much simpler and quicker in terms of communication rounds and processing complexity. Chandrasekharan et al. [9] focused on the issue of ensuring privacy when frequently mining in outsourcing transaction databases. They offer a novel strategy based on statistical discoveries on datasets to obtain k-support anonymity.

Baby et al. [10] proposed an effective homomorphic cryptographic technique for PPDM to secure data privacy. The main drawbacks of some of the currently available privacy-preserving technologies are their excessive computationally expensive and high communication overhead. They claim that their approaches guarantee absolute confidentiality and can withstand various attacks in the ARM process to some extent. H. Pang et al. [11] offer a new homomorphic cryptosystem that allows many cloud consumers to have unique public keys. They also offer a PPARM approach for data submitted from various parties. Their technique uses a transaction log description mechanism in archives for essential shopping centers. Underneath the cryptographic mining query on supermarket transactions, C. Ma et al. [12] provide an efficient approach for determining if an item set is frequent or not. They create a blocking method to boost mining efficiency. By dividing cryptographic transactions hooked on chunks and therefore only computing bilinear pairings on top of ciphertexts of portion blocks to a certain extent than all ciphertexts, this approach helps out to reduce the mining process' calculation cost.

Thakur et al. [13] have highlighted privacy-preserving mining on vertically partitioned datasets. In this situation, data owners want to investigate the affiliation regulations or standard item sets from such a shared dataset. They propose an effective homomorphic encryption system and a comfortable inspection service to ensure data security by suggesting a cloud-assisted frequent itemset mining approach to develop an ARM solution. The outcomes articulate outsourcing databases to allow various data owners to share their data while maintaining data privacy reliably. Compared to most existing arrangements, the outcomes expose minimal detail about the information. Given that all data and computing labor is an envoy to cloud servers, the information owner's beneficial resource intake may be minimal. For ARM over vertically partitioned databases, this work uses the D-Eclat algorithm, which would be more effective than that of the Eclat technique.

K. Agrawal et al. [14] present a cloud-assisted data analysis strategy for cloud services in a multi-party context with a little less basic data loss. They suggest a solution for mining association rules on outsourced data that consists of two moves: pre-processing and processing. At the client's end, the sanitizing stage completes. Each data owner uses the MD5 technique to anonymize each item and keeps records of each item and its forwarding rules value. Each data possessor delegates his protected dataset to a CSP, and all client IDs encipher with both the RC4 symmetric cryptographic algorithms (server). The mining process requires all processing to be on the server. They are integrating all of the encoded datasets into a single database. The shared database is accessed on the host using a decentralized network.

In cloud computing, H. Kim et al. [15] offer a PPARM technique for encrypted data. They use the Apriori algorithm with Elgamal cryptosystem to mine association rules, with no additional fraudulent transactions. As a result, the proposed approach may ensure data and query privacy while masking data frequency. S. Sharma et al. [16] suggested privacy-preserving graph spectral analysis techniques for cloud-based outsourced graphs. They envision a cloud-centric structure in which three parties collaborate: data providers, data owners, and cloud providers—using Adjacency matrices and Laplacian matrices to describe graphs. These matrices' elements are encrypted and supplied by remote contributors. After and then, the data owner engages with cloud-side programs to do spectrum analysis while safeguarding data protection from either the trustworthy or suspicious cloud service.

S. Qiu et al. [17] present a method for guaranteeing anonymity in pattern mining, in which information is collected and analyzed in encrypted form. On top of this architecture, they create three reliable, frequent pattern mining methods. They use two different homomorphic encryption algorithms and a safe and convenient comparing scheme to ensure data privacy and computing efficiency. The first protocol produces better mining efficiency, while the second protocol ensures greater privacy. As an information feature extraction technique, S. Priyadarsini et al. [18] employ ARM. For ARM, especially when employing the apriori formula. They updated the Apriori formula to make it more suitable for parallel computing. Lin Liu et al. [19] deal with the problem of data privacy for multiple parties. They devised a PP-ARM system that transfers data in a twin-cloud configuration. Based on the BCP cryptosystem, they create a set of encrypted chunks intended for ARM. Their method is dependent on the collection of advanced two-party dependable computational algorithms. They were able to attain a fair computing cost as well as a high level of secrecy.

M. Qaosar et al. [20] provided a system for performing a multi-party skyline query. Data anonymization, randomization, perturbation, and additive homomorphic encryption approaches are utilized. During the inquiry, the underlying protocols in the framework make sure that each collaborating participant recognizes its interring skyline objects without exposing them to everyone else. According to the extensive privacy and security evaluations, the approach can fulfill the desired processing goal without unauthorized disclosure. CPDE is a consent-based privacy-preserving decision tree evaluation technique proposed by L. Xue et al. [21]. In CPDE, the original decision tree assessment is performed in a confidential attempt to ensure model parameter confidentiality and user data privacy. As a result, all transactions can be done in the encrypted domain. By utilizing an additively homomorphic encryption elementary and a safe comparability protocol. The security examination demonstrates that CPDE satisfies the desired security features, while the throughput study shows that their approach is practical and acceptable for legitimate solutions.

## III. SYSTEM MODEL

The model divides into three components 1) Data owners (DO), 2) Intermediate cloud servers (ICS), and 3) Frequent itemsets computing cloud server (FCCS). The model starts with the data owner, who encrypts their data using the double encryption algorithm (DEA) to protect their data from the ICS and FCCS. After the encryption process, encrypted datasets are randomly partitioned and assigned to N clouds using the Transaction Splitter, a unique algorithm (TSA).

For organizing and maintaining encrypted information sent by DOs, intermediate cloud servers have a lot of storage space. Furthermore, ICS recognizes DOs' mining requests and works with FCCS to conduct privacy-preserving ARM. Each Intermediate cloud server collects encrypted sub-blocks (transactions) from various data owners, computes Local frequent itemsets using a customized Apriori (CA) algorithm, and delivers the encrypted Local frequent itemsets result to the FCCS to find the frequent global itemsets as shown in Fig. 2.

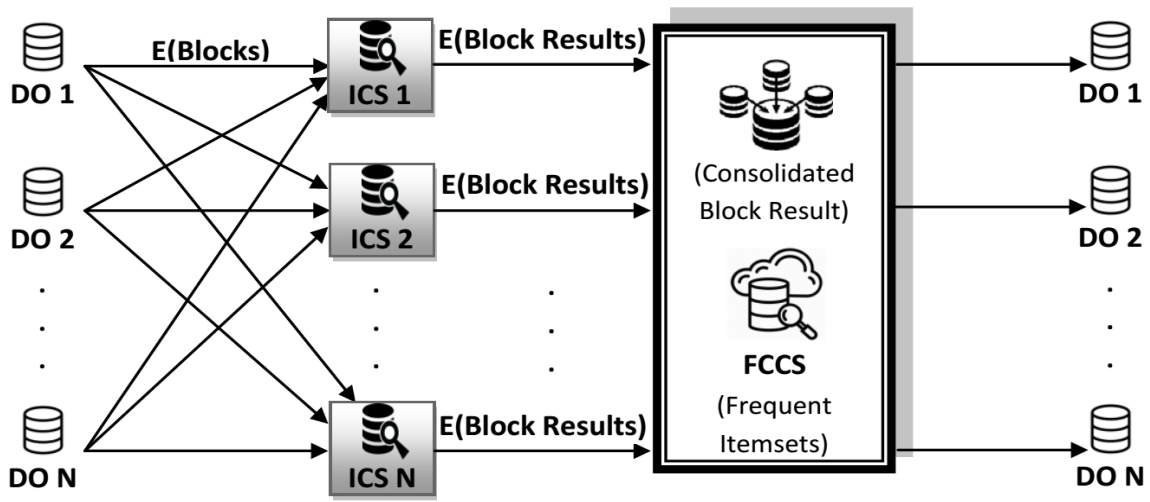

DO - Data Owner, ICS - Intermediate Cloud Server, FCCS - Frequent itemsets computing cloud server

**Fig. 2  System Model for Mining Privacy-Preserving Association Rules**

After receiving Local frequent item sets results from all Intermediate cloud servers, the Frequent item sets computing cloud server consolidates the Local frequent item sets results received from ICSs and computes the frequent global item sets, and distributes the results to all data owners as present in Fig. 1. The communication process for obtaining frequent global item sets will be different each time. Because the communication process depends on three factors: the data owners' key in a double encryption algorithm and the nature of the transaction splitter algorithm. Changes in the key will affect the double encryption algorithm, resulting in different outputs. The transaction splitter algorithm divides the data owners dataset based on the CLS randomly.

### A. Phase 1 – Secure Data Outsourcing

Data owners with minimal computational capabilities may perform some association rule mining operations in the cloud. This approach may expose data owners to the danger of sensitive personal information leakage. Data owners may encode original data before sharing it to protect outsourced data privacy. To maintain privacy, the data owner should convert the raw data thus Intermediate cloud servers cannot interpret or retrieve additional details than what the data owner has contributed.

Data owners encrypt their data sets using a double encryption algorithm to achieve a high degree of privacy during this stage. The preprocessing phase merely remains to be accomplished once per database [22]. After the encryption process, encrypted transactions are randomly partitioned (horizontally) and assigned to N clouds using the Transaction Splitter algorithm as shown in Fig. 3.

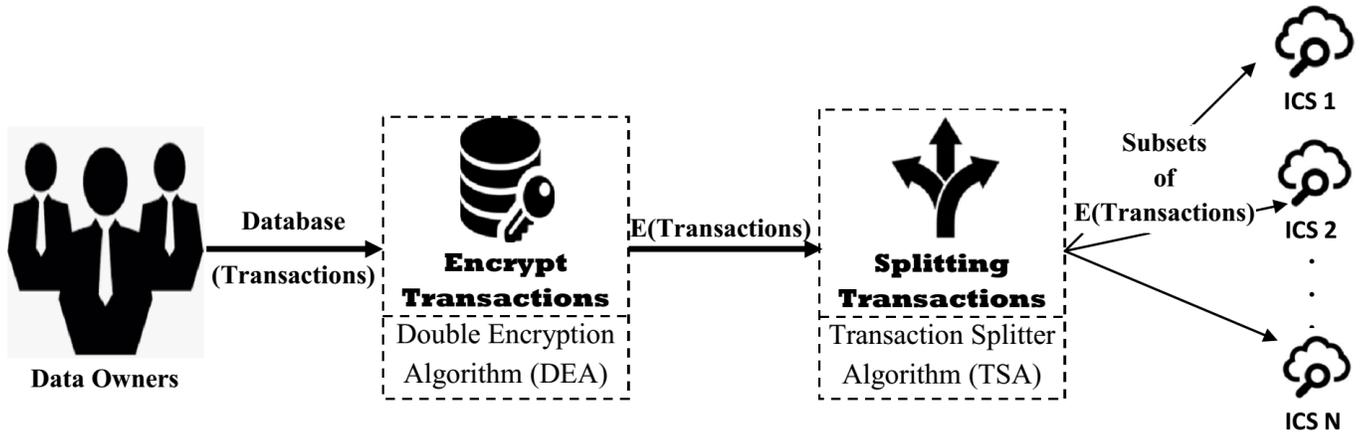

**Fig. 3 Phase I – Communication between the data owner and Intermediate cloud servers**

A transaction splitter algorithm is a horizontal data partition that encloses a subset of the whole information source and henceforth is conscientious for serving ICS's of the general responsibility. The transaction splitter algorithm level out an information source through horizontal fragmentation. A horizontal fragment of an information source is a subset of the tuples in that connection. A condition on at least one attribute of the relationship determines which tuples have a position with the horizontal fragment.

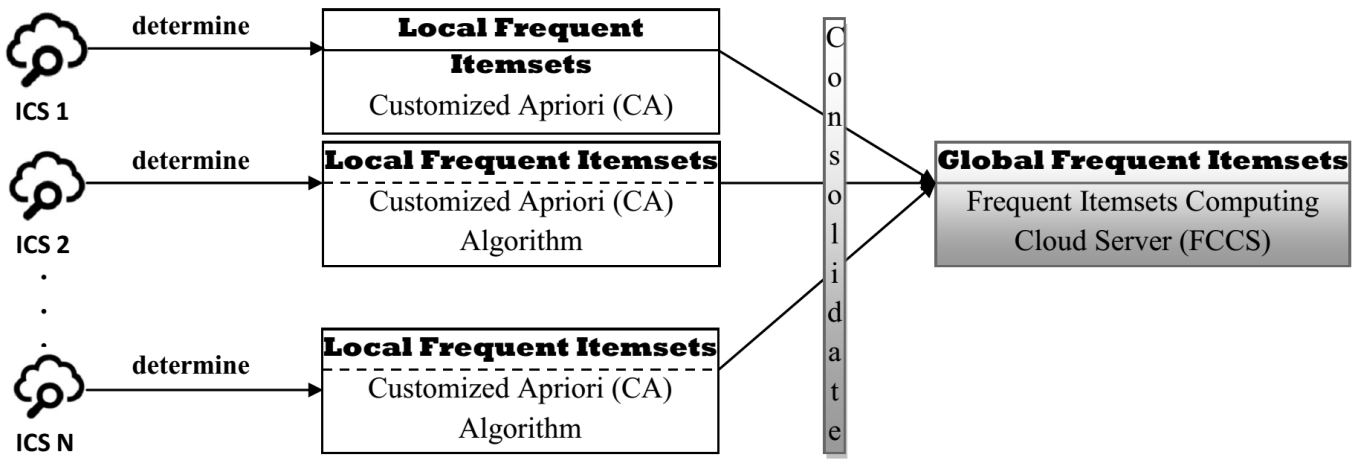

**Fig. 4 Phase II – Communication between Intermediate cloud servers and frequent itemsets computing cloud server**

**B. Phase 2 – Secure Computation of Frequent Itemsets**

The mining is carried out on the ICS's encrypted datasets. Let us imagine the data are the transactions that data owners have provided. Each transaction involves many goods, each of which is a one-of-a-kind product. Frequent pattern mining is to find item sets in many transactions. The amount of transactions containing X is the support of such an item set X, which is expressed as supp (X). The extraction solution is achieved by comparing supp (X) with a threshold min sup. If supp (X) is greater or equal to min sup, the result indicates that the item set X is common; X is infrequent, conversely.

During this stage, ICS determines the Local frequent item sets from the encrypted transactions received from diverse sources (Data Owners). To reduce the time complexity and maintain privacy, this paper proposes a customized Apriori algorithm to determine the Local frequent item sets, where ICS cannot apply frequency analysis attack. ICS's delivers the encrypted Local frequent item sets result to the Frequent item sets computing cloud server to find the frequent global item sets as shown in Fig. 4.

**C. Design Goals**

Proposed planned framework means to accomplish the accompanying five objectives:

- The framework permits data possessors to discard complicated computing processes.

- The framework permits parallel block computation, which diminishes the overall computation time.

- The framework gives the outrageous level of privacy to all data possessors without uncovering any data possessor's private information, intermediate outcomes, and Mined results (Association Rule).

- Customized Apriori algorithm, do not scan the entire database to count up the support in favor of each attribute.

- The model allows for more flexibility. For example, the framework will not collapse when many parties are involved.

**D. Double Encryption Algorithm**

Cryptography is crucial in ensuring the security of data exchanged over the internet. This paper aims to present a double encryption algorithm based on emerging combos strategies for data encryption. These strategies include suggesting a cryptographic approach [23]-[26] by combining conventional ciphers with sophisticated cipher methods to improve the confidentiality of the phase of the cryptographic function since cryptosystems based solely on conventional techniques are insecure. The unencrypted content to be transferred is encrypted using the Caesar Substitutions approach along with stream cipher [27]. As a result, the ICS's is ignorant of the encrypted content received.

**Double Encryption Steps:**

**Input:** Plain Text (PT)
**Output:** Double Encrypted Cipher Text (FCT)
**Encrypt - PlainText using Caesar cipher**
1) Convert the PlainText to ASCII value
2) Encrypt CT=PT+Key
3) where Key=5, PT is Plain Text, and CT is Cipher Text
4) Translate the CT to CT1

**Using the Stream cipher, encrypt the resultant CT again.**
5) Translate the CT1 to binary
6) Using CT=PT $\oplus$ Key, encipher the emerged CT
where the key of seven bits(binary)
7) Convert the seven-bit binary to ASCII value
8) Convert ASCII value to characters - final ciphertext (FCT)

**E. Transaction Splitter Algorithm**

The transaction splitter algorithm utilizes the data owner dataset as an input, including numerous transactions carried out by different clients. It utilizes an intelligent splitting approach to choose an arbitrary set of transactions depending on the number of intermediate cloud servers, preventing ICS from performing a frequency analysis attack [28] on the dataset they hold.

The Transaction splitter technique produced a random shuffling of transactions depending on N ICS. In O (1) time, it generates a random transaction sequence. The method winces from the most recent transaction id and replaces it with a chosen at random transaction id from the repository. Continue the cycle with the array of transaction ids spanning 1 to N-2 (length lowered by 1) until the criteria remove. The repeating process depends on the ICS count. The first cycle of the transaction splitter algorithm stops at the condition (Number of transactions/ICS). Once the condition stops, place the processed transaction id in a new list to be assigned for any one of the ICS. The second part of the transaction splitter algorithm identifies the difference between the new list and the original transaction id list [29]. The difference will be placed in a new list and assigned as the original list. The loop executes until the condition N(ICS) reaches.

**Transaction Splitter Algorithm**

**Input:** A List, Data transaction T, Count of ICS
**Output:** A List, L of N blocks based on ICS
Step 1: To Generate a random choice of Transaction ID
        srand(time(NULL))
        index = rand() % T // T - number of Transaction ID
Step 2: Get random Transaction ID from the Transaction
        vector
          num = v[index]
Step 3: Remove the Transaction ID from the Transaction
        vector
        swap(v[index], v[T - 1]);
        v.pop_back();

Step 4: Function to generate n non-repeating random
    Transaction ID
      for ID starts from 0 and i < ( T/ICS)
      v[i] = i + 1;
Step 5: get a random Transaction ID from the Transaction
    vector
      while (v.size())
    getNum(v) // v= List-i(ICS1) of random Transaction ID
      increment i

**To find the Lists of Transaction apart from List-x(ICS-x)**

Step 6: Two Lists say List and List-x.
Step 7: Pick one Transaction ID from List and compare it
    with all the Transaction ID of the List x
Step 8: If the Transaction ID of List exists in List x,
    discard that Transaction ID and
    pick up the next Transaction ID of List and
    repeat step 7.
Step 9: If the Transaction ID of List does not exist in List x,
    add that Transaction ID in List y.
    Before adding that Transaction ID to List y,
    ensure that Transaction ID does not already exist in
    List y
Step 10: Repeat steps 7 to 9 until all the Transaction IDs of
    List are compared
Step 11: list-y contains all Transaction IDs that represent the
    difference between List and list-x
where List-Main List of Transaction IDs, List-x for ICS-x.
Step 12: Repeat step 1 to 11 until Transaction IDs grouped as
    List and distributed evenly to all ICS.

The transaction splitter algorithm has the feature in which each ICS does have an equal chance of picking as the final transaction id inside the sequence of transaction id, which is 1/N. As a result, obtaining a well-shuffled sequence of transaction ids after the partitioning operation. Consequently, the LN subsidiary list will randomly arrive in an ICS list.

### F. Customized Apriori Algorithm

Multiple searches across the entire data were a significant weakness in the Apriori algorithm. It necessitated a significant amount of room and time. According to the change in this article, no need to run the entire database to determine the support for each characteristic. This can be done by keeping track of the minimal support count and contrasting it to the support of each characteristic. It tracks an attribute's support until it achieves the minimum support value.

Support for an attribute does not have to be recognized further than that. The use of a STOP variable in the algorithm allows for this provision. The sequence is interrupted and indeed, the result for support is logged as soon as STOP updates its value.

**Algorithm - Customized Apriori**

**Input :** Transactions TD; min_supp, Min_sup Limit
**Output :** Ls, frequent item sets in TD

1) Ls(1)= locate_freq _1-item sets(TD);
2) For each one transaction tr in TD
3) cou_i= cou_i (tr); // where cou_i is item count
4) For (a=2; Ls(a-1)!=null; a++)
    C(i) = apri_gene(Ls(a-1, min_supp);
    STOP=1;
5) For each one transaction tr fit in to TD
    cou_i >=a
6) If (STOP==1)
    c=part(C(a),tr);
    c.cou++;
    If (c.cou==min_supp)
    STOP=0;
7) If (STOP==0)
    go out from loop
Ls(a)={c.cou=min_sup}
return Ls=U(a) Ls(a);

### G. Computing Global Block Result- Frequent itemsets computing cloud server

The ICS results are communicated to the FCCS once each ICS has computed frequent itemset for the specified block. From the computed block result received from all the ICS, FCCS will aggregate the frequent itemset and identify the frequent global itemset and the Association rule. FCCS will display the frequent global itemset to all other data owners through the portal.

$$\text{Frequent Itemsets (FI)} = \sum_{f=1}^{N} \sum_{ICS=1}^{n} \left( (FI)_f \right)_{ICS} \qquad (1)$$

## IV. PERFORMANCE EVALUATION

In this part, by running several tests on real-world data sets to see how well the unique algorithms perform in terms of the computational complexity of the proposed customized ARM and frequent itemset mining methods. First, by using double crypto to encrypt [30] data using a 32-bit modulus form and this paper builds the double cryptosystem code with Python's math and random libraries. Then, as a pattern, this paper uses one of [28]'s solutions and common non-privacy preserving tactics. The proposed technique achieves a high level of privacy with no enlightening any other information as regards the users' data. Other techniques, on the other hand, achieve lower degrees of security. Because they are now more effective ways accessible, traditional non-privacy algorithms employ as baselines.

#### Table 1. Experimental settings

| CPU | Intel(R) Core i5-2410M @ 2.30 GHz |
|---|---|
| Software | Windows 7 64-bit and Apache NetBeans |
| Memory | 8.00 GB |
| Data | https://data.world/datasets/health |
| Data bit length | < 72 bits |
| $\lambda$ (security Variable) | 80 |
| \|ra\| for $Eb_s$, $Eb_e$ and ERVs | = 40 bits = $\lambda/2$ |
| \|ra\| for $Eb_z$ | 64   its |

### A. Comparing with Traditional Non-Privacy Algorithms

Using the running time, calculate the computational complexity. To highlight the usefulness of this work, examine these methods to currently available non-privacy preserving techniques. This paper assesses the proposed concepts and traditional algorithms using medical scrutinizing datasets obtained from "https://data.world/datasets/health," to assess proposed concepts and traditional algorithms. This paper randomly divides all resources into t data sets to mimic t data owners.

#### Table 2. Runtime comparison (t = 4 and k = 10)

| Number of Transactions | DO | Cloud | Apriori | Eclat | Fp-growth |
|---|---|---|---|---|---|
| 500 | 6 | 39 | 27 | 11 | 9 |
| 1000 | 6 | 40 | 25 | 12 | 5 |
| 3000 | 5 | 38 | 19 | 9 | 8 |
| 10000 | 7 | 41 | 23 | 8 | 4 |
| 15000 | 7 | 33 | 17 | 9 | 7 |

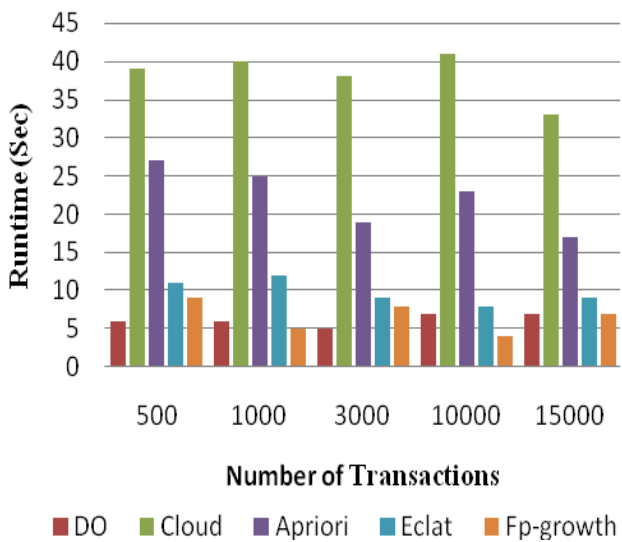

**Fig. 5 Runtime comparison (t = 4 and k = 10) with traditional non privacy Algorithms**

This paper utilizes NetBeans for the proposed work and the research employs a JAVA completion of the FP-growth, Apriori, and Eclat algorithms. In all studies, ten laptops taking the role of cloud and data owners had the same software and hardware settings to ensure a natural correlation. Employing four laptops as data owners, which decode the data and send it to the cloud, is unusual—the remaining items as the cloud and Evaluator, respectively.

#### Table 3. Runtime comparison (diverse data owner)

| Number of Transactions | t=4 | t=7 | t=10 |
|---|---|---|---|
| Cloud Ts=500 | 25 | 30 | 37 |
| Cloud Ts=5000 | 24 | 26 | 31 |
| DO Ts=500 | 6 | 4 | 3 |
| DO Ts=5000 | 5 | 4 | 4 |

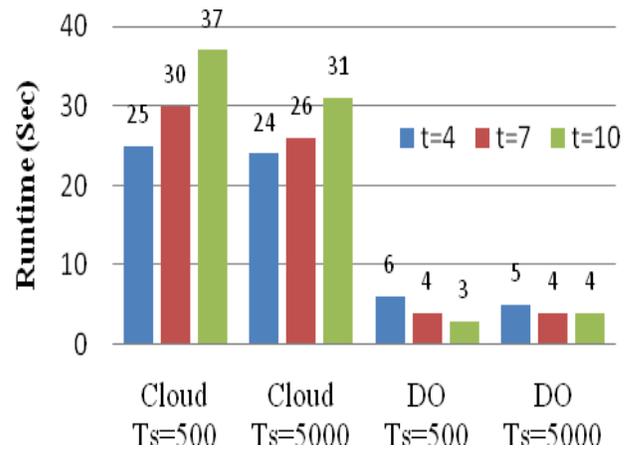

**Fig. 6 Runtime under diverse data owner count t (The value of k set to 12)**

Table 2-5 shows the runtime comparison of ARM. The experimental parameters for proposed solutions are in Tab. 1. Figures 5, 6, 7, and 8 show the results (running time) of ARM. The runtime is into two parts: cloud end and data owner end. This paper shows that proposed solutions have a one-order-of-magnitude run time longer than the most desirable non-PPDM algorithms based on the data analysis employing multiple parameters (t, k, and c) and data sources. Because ICS does both information and processing work, the data owner's resource use is less. Proposed methods are just as successful as the most cutting-edge low-privacy methods. In most circumstances, the CS takes one order of magnitude longer to run than the old approach, and the data owner takes one order of magnitude less.

**Table 4. Runtime comparison (Itemset)**

| Number of Transactions | k=8 | k=12 | k=14 |
|---|---|---|---|
| Cloud Ts=500 | 13 | 16 | 19 |
| Cloud Ts=5000 | 31 | 33 | 33 |
| DO Ts=500 | 8 | 7 | 6 |
| DO Ts=5000 | 7 | 7 | 4 |

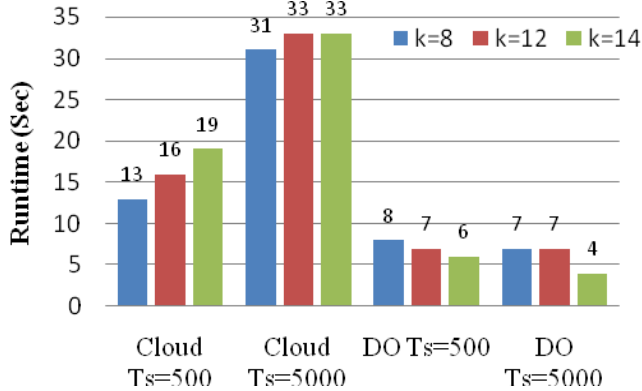

**Fig. 7 Runtime under diverse Itemset k (t is set to 4)**

Runtime changes with increasing k and t upsides in Figs. 6 and 7. With the criteria k and t, the cloud runtime increases. This paper measures the runtime of cloud increments when k is set to 12 and the criterion t scaling from 4 to 10. In both conventions, the data owner's runtime decreases.

**Table 5. Runtime comparison**

| Number of Transactions | c=4 | c=5 | c=6 |
|---|---|---|---|
| Cloud Ts=500 | 9 | 9 | 9 |
| Cloud Ts=5000 | 10 | 9 | 10 |
| DO Ts=500 | 4 | 4 | 5 |
| DO Ts=5000 | 4 | 4 | 4 |

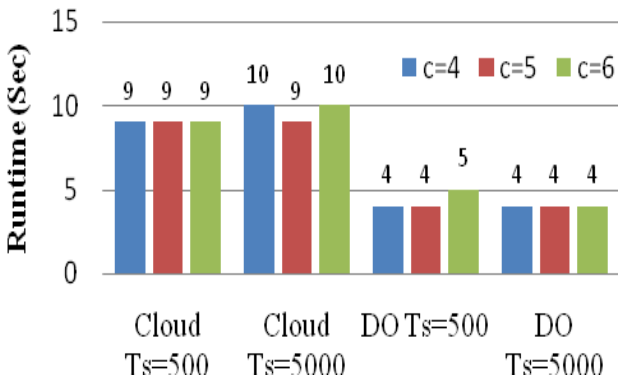

**Fig. 8 Runtime under multiple cloud c (t is fixed to 4 and k is fixed to 12)**

Whenever the volume of DO rises to four and the criterion k increases from eight to fourteen, This paper discovers that the cloud runtime increases, and the data owner runtime changes only slightly. By setting the criterion t to three and the criterion k to ten, runtime decreases as ICS increases. The data owner's runtime changes as the number of ICS increases. If t grows, the data owner's runtime declines; when a similar joint database divides into additional DO, each DO's dataset shrinks. As a result, it takes less time to preprocess a smaller dataset. As a result, the running time of data owners does not rise as k increases. Using the proposed approaches, expanding k and t results in a longer runtime on the cloud side but no increase in the data owner's runtime.

**B. Comparing with Privacy Algorithms**

Using the same arrangement of hardware, software, and datasets to execute contrasts with non-privacy Algorithms.

**Table 6. Runtime comparison**

| System | Time in ms |
|---|---|
| Existing system (PCML) without horizontal segmenting with 10000 transactions | 63200 |
| A proposed system with horizontal segmenting with 10000 transactions | 41300 |

The computation cost of the proposed solution is a lot of substandard identified with the disinfected dataset, Transaction Splitter algorithm, and outwork customized ARM. One of the most incredible existing privacy-preserving solutions that do not release sensitive data of the raw information is [28]'s frequent itemset mining utilizing privacy-preserving collaborative model learning (PCML) scheme.

**Table 7. Computation cost comparison**

| Number of Transactions | DO | Cloud | PCML |
|---|---|---|---|
| 500 | 5 | 30 | 61 |
| 1000 | 6 | 41 | 66 |
| 10000 | 6 | 35 | 71 |
| 20000 | 7 | 41 | 67 |
| 25000 | 6 | 23 | 73 |

Table 6 and 7 depict the time requirements for executing the existing and proposed frameworks, as determined by the mathematical model. The execution time of this solution is more than the proposed framework because of its costly operations. Fig. 9 analyzes the computation cost of the proposed methodology with a PCML scheme by changing the number of transactions in addition to the range of criteria (t, k, c) to track down the frequent item sets. This paper sets the criterion value as DO=3, k=6, and c=4.

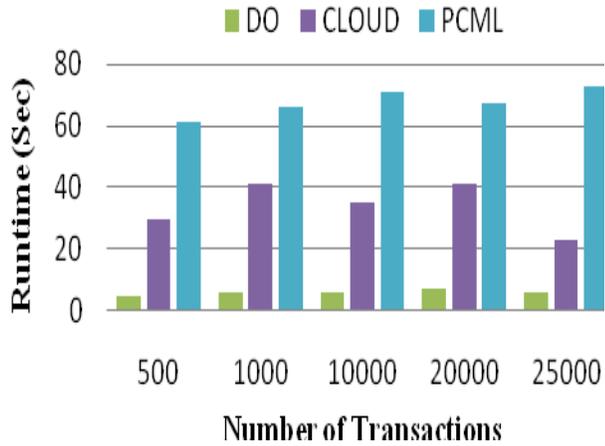

**Fig. 9 Computation Cost Analysis with privacy Algorithm**

This paper guarantees that the proposed design and execution are versatile when the data size increments based on the outcome. This paper likewise sees that the runtime of proposed protocols is practically comparative, and the runtime of proposed protocols is more diminutive in contrast with the PCML scheme.

## V. CONCLUSION

In data mining, association rule mining is a common practice intended for uncovering meaningful relationships between elements in massive databases. The goal is to use numerous possible methods to find strong rules uncovered in databases. In collaborative ARM, privacy protection is critical when data is outsourced. This paper addresses the topic of safe ARM when transactions disperse over cloud servers in this work. This paper provides three unique algorithms that can reach excellent data utility, confidentiality, and time consumption while also reaching higher data utility and confidentiality. This paper proposes a unique double encryption and Transaction Splitter approach to alter the database to optimize the utility and privacy tradeoff in the preparation phase.

This paper presents a customized Apriori approach for the mining process, in which exploring the entire database to assess the support for each attribute is not needed. According to mathematical theory and research data, the proposed technique can output correct mining results and has a substantially quicker operating time. As a result, it is much more effective than earlier approaches with almost the same degree of security, processing mining on an encrypted mining request. In the future, the focal point is to increase the reliability and performance of pattern mining taking place in large-scale sparse data, as well as apply this approach to additional contexts.